\documentclass[letterpaper, 10 pt, conference]{ieeeconf}

\IEEEoverridecommandlockouts
\usepackage[utf8]{inputenc}
\usepackage[T1]{fontenc}
\usepackage{cite}
\usepackage{amsmath,amssymb,amsfonts}
\usepackage{graphicx}
\usepackage{textcomp}
\usepackage{xcolor}
\usepackage{mathtools}

\usepackage{bm}
\usepackage{float}
\usepackage{hyperref}
\usepackage{amsthm}
\newenvironment{compactproof}{%
  \begin{proof}\begingroup
    \setlength{\abovedisplayskip}{2pt}
    \setlength{\belowdisplayskip}{2pt}
    \setlength{\abovedisplayshortskip}{2pt}
    \setlength{\belowdisplayshortskip}{2pt}
    \linespread{0.8}\selectfont
}{%
    \endgroup\end{proof}
}

\usepackage{etoolbox}

\newtheorem{theorem}{Theorem}

\theoremstyle{remark}
\newtheorem{remark}{Remark}

\newtheorem{definition}{Definition}
\newtheorem{proposition}{Proposition}

\DeclareMathAlphabet{\mathcal}{OMS}{cmsy}{m}{n}

\makeatletter
\renewcommand\paragraph{\@startsection{paragraph}{4}{\z@}%
  {1.0ex \@plus 0.3ex \@minus 0.2ex}%
  {0.4em}%
  {\normalfont\normalsize}}
\makeatother

\def\BibTeX{{\rm B\kern-.05em{\sc i\kern-.025em b}\kern-.08em
    T\kern-.1667em\lower.7ex\hbox{E}\kern-.125emX}}

\graphicspath{{image/}}

\begin{document}

\title{Bridging Data-Driven Reachability Analysis and Statistical Estimation via Constrained Matrix Convex Generators%
\thanks{P.~Xie, Z.~Zhang and A.~Alanwar are with the TUM School of Computation,
Information and Technology, Department of Computer Engineering,
Technical University of Munich, 74076 Heilbronn, Germany.
\texttt{(e-mail: p.xie@tum.de, zhangzhenzhang@tum.de, alanwar@tum.de)}}%
\thanks{Rolf Findeisen is with the Control and Cyber-Physical Systems Laboratory (CCPS), Technical University of Darmstadt, 64283 Darmstadt, Germany. \texttt{(e-mail: rolf.findeisen@iat.tu-darmstadt.de)}}%
}

\author{Peng Xie, Zhen Zhang, Rolf Findeisen, Amr Alanwar}

\maketitle

\begin{abstract}
Data-driven reachability analysis enables safety verification when first-principles models are unavailable. This requires constructing sets of system models consistent with measured trajectories and noise assumptions. Existing approaches rely on zonotopic or box-based approximations, which do not fit the geometry of common noise distributions such as Gaussian disturbances and can lead to significant conservatism, especially in high-dimensional settings. This paper builds on ellipsotope-based representations to introduce mixed-norm uncertainty sets for data-driven reachability. The highest-density region defines the exact minimum-volume noise confidence set, while Constrained Convex Generators (CCG) and their matrix counterpart (CMCG) provide compatible geometric representations at the noise and parameter level. We show that the resulting CMCG coincides with the maximum-likelihood confidence ellipsoid for Gaussian disturbances, while remaining strictly tighter than constrained matrix zonotopes for mixed bounded-Gaussian noise. For non-convex noise distributions such as Gaussian mixtures, a minimum-volume enclosing ellipsoid provides a tractable convex surrogate. We further prove containment of the CMCG × CCG product and bound the conservatism of the Gaussian–Gaussian interaction. Numerical examples demonstrate substantially tighter reachable sets compared to box-based approximations of Gaussian disturbances. These results enable less conservative safety verification and improve the accuracy of uncertainty-aware control design.

\end{abstract}

\section{Introduction}\label{sec:intro}

Reachability analysis computes the set of all states a dynamical system can reach under all admissible inputs and disturbances, a fundamental tool for safety verification~\cite{althoff2010reachability,girard2005reachability,kuhn1998rigorously}.  When a first-principles model is unavailable, data-driven methods compute reachable sets directly from measured input-state trajectories.  A central ingredient is the model set of all system matrices consistent with the data and a noise assumption.  Under bounded noise, the model set can be represented as a constrained matrix zonotope (CMZ), building on zonotopic uncertainty representations widely used in set-based estimation and fault diagnosis~\cite{scott2013input}, and propagated forward in time~\cite{amr23reachable,alanwar2022data,alanwar2022enhancing}.

Existing probabilistic zonotope methods~\cite{althoff2015introduction} truncate Gaussian confidence regions with $\infty$-norm boxes, though the natural geometry is the $2$-norm ball.  In dimension $q$, this inflates the confidence-region volume by $2^q / V_q$ ($6\times$ for $q\!=\!5$, $310\times$ for $q\!=\!10$).  This paper replaces the $\infty$-norm truncation by the mixed-$p$ geometry of ellipsotopes~\cite{kousik2023ellipsotopes}, and carries this correction through model-set construction and propagation.  The Highest Density Region (HDR)~\cite{hyndman1996hdrgraph} gives the statistically exact noise confidence region; the Constrained Convex Generators (CCG) representation provides the set calculus for pullback and propagation.  For non-convex HDRs from Gaussian-mixture noise, we include a preliminary treatment based on the minimum-volume enclosing ellipsoid (MVEE).

The paper makes three contributions.  First, it shows how mixed-$p$ CCG/CMCG sets can be used systematically in data-driven reachability for bounded, Gaussian, mixed bounded-Gaussian, and (via an MVEE surrogate) Gaussian-mixture noise.  Second, it proves a pullback theorem from noise-level CCG to parameter-level CMCG and shows that, by exploiting the orthogonal projection independence of Gaussian noise, the CMCG coincides with the MLE confidence ellipsoid ($\text{CMCG} = \text{MLE} \subset \text{CMZ}$).  Third, it proves containment of the CMCG $\times$ CCG product and bounds the Gaussian$\times$Gaussian truncation conservatism.

Constrained zonotopes were introduced in~\cite{scott2016constrained}; ~\cite{amr23reachable} extended the idea to data-driven reachability with matrix zonotopes.  Probabilistic zonotopes~\cite{althoff2015introduction} combine bounded and Gaussian uncertainty but truncate the Gaussian part with $\infty$-norm boxes.  \cite{kousik2023ellipsotopes} introduced ellipsotopes, unifying ellipsoids and zonotopes; CCG extends this to mixed $p$-norms.  The work in~\cite{csaji2012sps} developed the Sign-Perturbed Sums method for exact finite-sample confidence regions.

The results of this paper establish a principled connection between statistical estimation and data-driven reachability by aligning uncertainty representations with the underlying noise geometry. In particular, the proposed CMCG representation recovers the maximum-likelihood confidence set for Gaussian disturbances while avoiding the conservatism induced by box-based approximations, and extends naturally to mixed bounded and stochastic uncertainty. This enables substantially tighter reachable sets and provides a foundation for less conservative safety verification and uncertainty-aware control design in data-driven settings. 

The remainder of the paper introduces the proposed set representations, derives the corresponding parameter sets via pullback, and develops tractable propagation schemes together with numerical validation.

\section{Preliminaries and Problem Statement}\label{sec:preliminaries}

Matrices are denoted by capitals ($A$, $B$), vectors by lowercase ($x$, $c$), sets by calligraphic letters ($\mathcal{Z}$, $\mathcal{M}$).  The identity matrix is $I$, $\mathbb{R}^n$ is $n$-dimensional Euclidean space, time indices are subscripts ($x_k$), and $M^\dagger$ denotes the Moore--Penrose pseudoinverse.

\subsection{Zonotope and matrix zonotope}

\begin{definition}[Zonotope \cite{girard2005reachability}]
A zonotope $\mathcal{Z} \subset \mathbb{R}^n$ with center $c \in \mathbb{R}^n$ and generator matrix $G \in \mathbb{R}^{n \times \gamma}$ is the set
\begin{equation}
\mathcal{Z} = \langle c, G \rangle
:= \Big\{c + G\beta \ \Big| \ \|\beta\|_\infty \le 1 \Big\}.
\label{eq:zonotope_def}
\end{equation}
\end{definition}

\begin{definition}[Matrix zonotope \cite{althoff2010reachability}]
A matrix zonotope $\mathcal{M} \subset \mathbb{R}^{n \times p}$ with center $C \in \mathbb{R}^{n \times p}$ and generators $G^{(i)} \in \mathbb{R}^{n \times p}$, $i = 1,\ldots,\gamma$, is the set
\begin{equation*}
\mathcal{M}
\!=\! \big\langle C, G^{(1)}\!, \ldots, G^{(\gamma)} \big\rangle
\!:=\! \Big\{C \!+\! \textstyle\sum_{i=1}^{\gamma} \beta_i G^{(i)}
\;\Big|\; \|\beta\|_\infty \!\le\! 1 \Big\}.
\label{eq:mz_def}
\end{equation*}
\end{definition}

Zonotopes are closed under linear maps and Minkowski sums: for $R \in \mathbb{R}^{m \times n}$ and two zonotopes $\mathcal{Z}_1 = \langle c_1, G_1 \rangle$, $\mathcal{Z}_2 = \langle c_2, G_2 \rangle$,
\begin{equation*}
R\,\mathcal{Z}_1 = \langle Rc_1, RG_1 \rangle,
\qquad
\mathcal{Z}_1 \oplus \mathcal{Z}_2
= \big\langle c_1+c_2,\
[G_1 \ G_2] \big\rangle.
\label{eq:zono_ops}
\end{equation*}

\subsection{Constrained Convex Generators (CCG)}

Kousik et al.~\cite{kousik2023ellipsotopes} introduced ellipsotopes, which partition the coefficient vector into index groups each constrained by a $2$-norm, and noted that other $p$-norms could be assigned per group~\cite[Remark~6]{kousik2023ellipsotopes}. We adopt this mixed-$p$ extension and call the resulting sets Constrained Convex Generators (CCG), reserving ``ellipsotope'' for the case $p_k = 2$ for all groups.

\begin{definition}[Constrained Convex Generators (CCG)~\cite{kousik2023ellipsotopes}]
\label{def:ccg}
A CCG set $\mathcal{E} \subset \mathbb{R}^n$ is defined as
\begin{equation*}
\mathcal{E}
= \left\{c + G\beta
\;\middle|\;
\|\beta_{\mathcal{I}_k}\|_{p_k} \le 1\ \forall\, k=1,\ldots,K,\
A\beta = b
\right\},
\label{eq:ccg_def}
\end{equation*}
where $c \in \mathbb{R}^n$ is the center, $G \in \mathbb{R}^{n \times m}$ is the generator matrix, $\{\mathcal{I}_k\}_{k=1}^{K}$ are disjoint index sets partitioning the coefficients $\beta$, each with its own norm $p_k$, and $A\beta = b$ are optional linear equality constraints.
\end{definition}

Special cases: $p_k = 2$ gives an ellipsotope~\cite{kousik2023ellipsotopes}; all $p_k = \infty$ with singleton index sets and no constraints gives a zonotope; adding linear constraints gives a constrained zonotope~\cite{scott2016constrained}; different $p_k$ values produce a mixed-index CCG.

\begin{definition}[Constrained Matrix Convex Generators (CMCG)]
\label{def:cmg}
A CMCG $\mathcal{N} \subset \mathbb{R}^{n \times p}$ is defined as
\begin{align}
\mathcal{N}
\!:=\! \Big\{
C\! + \!\textstyle\sum_{k=1}^{\gamma} \beta_k G^{(k)}
\;\Big|\nonumber 
\|\beta_{\mathcal{I}_j}\|_{p_j} \!\le\! 1\ \forall\, j,\
\!\!\textstyle\sum_k \beta_k A^{(k)}\!\! =\! B
\Big\},
\label{eq:cmg_def}
\end{align}
where $C$ is the center matrix, $G^{(k)}$ are generator matrices, and $A^{(k)}$, $B$ define the linear equality constraints.
\end{definition}

The CMCG is the matrix form of the CCG, used to represent parameter sets.  When all norms are $p_j = \infty$ with singleton index sets, the CMCG reduces to a constrained matrix zonotope (CMZ)~\cite{amr23reachable}.

\subsection{Probabilistic zonotope and probabilistic matrix zonotope}
\label{sec:probZ}

\begin{definition}[Probabilistic zonotope~\cite{althoff2010reachability}]
\label{def:probz}
A probabilistic zonotope $\mathcal{Z}_p \subset \mathbb{R}^n$ with center $c \in \mathbb{R}^n$, bounded generators $G_b \in \mathbb{R}^{n \times \gamma_b}$, and Gaussian generators $G_g \in \mathbb{R}^{n \times \gamma_g}$ is the set
\begin{equation}
\mathcal{Z}_p = \Big\{ c + G_b \beta + G_g \xi \;\Big|\; \|\beta\|_\infty \le 1,\; \xi \sim \mathcal{N}(0, I_{\gamma_g}) \Big\}.
\label{eq:probz_def}
\end{equation}
\end{definition}

\begin{definition}[Probabilistic matrix zonotope~\cite{althoff2010reachability}]
\label{def:probmz}
A probabilistic matrix zonotope $\mathcal{M}_p \subset \mathbb{R}^{n \times p}$ with center $C \in \mathbb{R}^{n \times p}$, bounded generators $G_b^{(i)} \in \mathbb{R}^{n \times p}$, $i = 1,\ldots,\gamma_b$, and Gaussian generators $G_g^{(j)} \in \mathbb{R}^{n \times p}$, $j = 1,\ldots,\gamma_g$, is the set
\begin{multline}
\mathcal{M}_p = \Big\{ C + \textstyle\sum_{i=1}^{\gamma_b} \beta_i G_b^{(i)} + \textstyle\sum_{j=1}^{\gamma_g} \xi_j G_g^{(j)} \;\Big|\;\\
\|\beta\|_\infty \le 1,\; \xi \sim \mathcal{N}(0, I_{\gamma_g}) \Big\}.
\label{eq:probmz_def}
\end{multline}
\end{definition}

\begin{proposition}[Confidence truncation: from probabilistic zonotope to CCG]
\label{prop:probz_to_ccg}
Let $\mathcal{Z}_p$ be a probabilistic zonotope (Definition~\ref{def:probz}) with Gaussian generators $G_g \in \mathbb{R}^{n \times \gamma_g}$, and let $1-\alpha$ be a prescribed confidence level.  Define the truncation radius
\begin{equation}
\rho := \sqrt{\chi^2_{\gamma_g,\,1-\alpha}},
\label{eq:truncation_radius}
\end{equation}
where $\chi^2_{\gamma_g,1-\alpha}$ denotes the $(1-\alpha)$-quantile of the chi-squared distribution with $\gamma_g$ degrees of freedom.  Then the $(1-\alpha)$-confidence truncation of $\mathcal{Z}_p$ is the CCG
\begin{multline}
\mathcal{Z}_p^{1-\alpha} = \Big\{ c + G_b \beta^{(b)} + \rho\, G_g \beta^{(g)} \;\Big|\;\\
\|\beta^{(b)}\|_\infty \le 1,\; \|\beta^{(g)}\|_2 \le 1 \Big\},
\label{eq:probz_truncated}
\end{multline}
with index groups $\mathcal{I}_b$ for the bounded coefficients ($p_b = \infty$) and $\mathcal{I}_g$ for the Gaussian coefficients ($p_g = 2$).  The same construction applied to a probabilistic matrix zonotope yields a CMCG.
\end{proposition}

\begin{compactproof}
Since $\xi \sim \mathcal{N}(0, I_{\gamma_g})$, $\|\xi\|_2^2 \sim \chi^2_{\gamma_g}$, so $\Pr\{\|\xi\|_2 \le \rho\} = 1 - \alpha$.  Substituting $\beta^{(g)} := \xi / \rho$ maps $\{\|\xi\|_2 \le \rho\}$ to $\{\|\beta^{(g)}\|_2 \le 1\}$ with $G_g \xi = \rho\, G_g \beta^{(g)}$.  The resulting set~\eqref{eq:probz_truncated} has exactly the CCG structure of Definition~\ref{def:ccg} with $p_b = \infty$ and $p_g = 2$.
\end{compactproof}

\begin{remark}[Norm mismatch in prior probabilistic zonotope approaches]
\label{rem:norm_mismatch}
Prior work~\cite{althoff2015introduction} truncates with $\|\xi\|_\infty \le m$, whereas the true $(1-\alpha)$ confidence region is $\|\xi\|_2 \le \sqrt{\chi^2_{q,1-\alpha}}$. The box inflates the volume by $2^q / V_q$ ($V_q$ = unit $q$-ball volume):

\vspace{2pt}
\begin{center}
\begin{tabular}{c c c}
\hline
$q$ & $2^q / V_q$ & over-approx.\ \\
\hline
$2$ & $1.27$ & $27\%$ \\
$5$ & $6.08$ & $508\%$ \\
$10$ & $310$ & $31{,}000\%$ \\
\hline
\end{tabular}
\end{center}
\vspace{2pt}

The CCG avoids this inflation by using the correct $2$-norm for Gaussian generators.
\end{remark}

\begin{figure}[t]
    \centering
    \includegraphics[width=0.92\columnwidth]{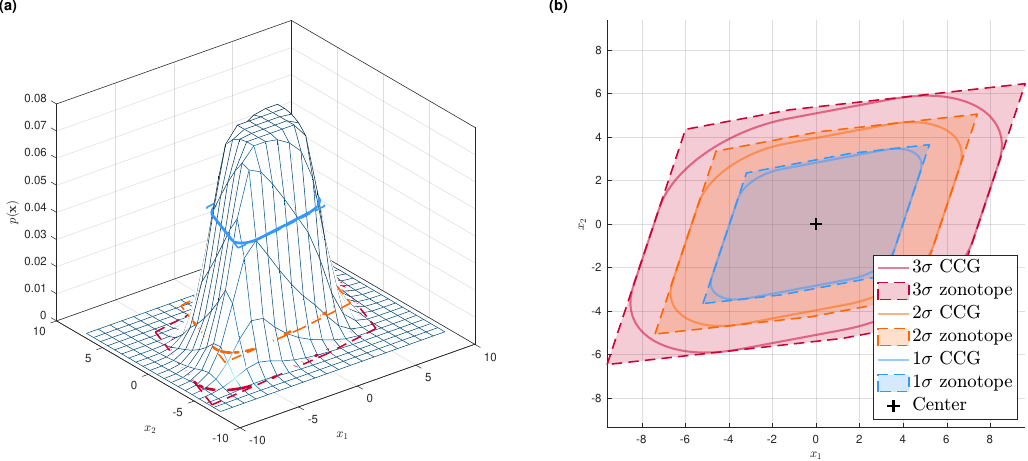}
    \caption{Mixed bounded-Gaussian truncation.  (a)~3D density surface.  (b)~$m\sigma$ level sets: CCG (solid) vs.\ probabilistic zonotope (dashed).  The CCG uses a $2$-norm ball for the Gaussian part, avoiding the box over-approximation.}
    \label{fig:truncation_mixed}
\end{figure}

Figure~\ref{fig:truncation_mixed} illustrates this for the mixed bounded-Gaussian case: the CCG (solid) uses a $2$-norm ball for the Gaussian part, while the probabilistic zonotope (dashed) over-approximates it with a box.

\subsection{Highest Density Region (HDR)}
\label{sec:hdr_def}

\begin{definition}[Highest Density Region~\cite{hyndman1996hdrgraph}]
\label{def:hdr}
Given a density $f_W$ on $\mathbb{R}^q$, the $(1-\alpha)$ highest density region (HDR) is defined as
\begin{equation}
\mathcal{H}_{W,1-\alpha} := \{w \in \mathbb{R}^q : f_W(w) \geq \tau_\alpha\},
\label{eq:hdr_def}
\end{equation}
where $\tau_\alpha$ is the largest threshold such that $\Pr\{W \in \mathcal{H}_{W,1-\alpha}\} \geq 1-\alpha$.
\end{definition}

\begin{remark}[Properties of the HDR]
The HDR is the smallest-volume set with coverage $1-\alpha$~\cite{hyndman1996hdrgraph}. For bounded and Gaussian noise it is convex, whereas for Gaussian-mixture noise it can be non-convex and disconnected.
\end{remark}

\subsection{Problem statement}

Consider a discrete-time linear time-invariant system
\begin{equation}
x_{k+1} = Ax_k + Bu_k + w_k,
\label{eq:lti_system}
\end{equation}
where $x_k \in \mathbb{R}^n$, $u_k \in \mathbb{R}^m$, and $w_k \in \mathbb{R}^n$ are the state, input, and process disturbance.  The matrices $A \in \mathbb{R}^{n \times n}$, $B \in \mathbb{R}^{n \times m}$ are unknown.  We assume access to a trajectory $\{(u_0, x_0), \ldots, (u_{T-1}, x_{T-1}), x_T\}$.

Given initial set $\mathcal{X}_0 \subset \mathbb{R}^n$ and input set $\mathcal{U} \subset \mathbb{R}^m$, the reachability problem is to enclose all states reachable at time $k$ under all data-consistent system matrices and admissible disturbances. We assume that the noise density $f_W$ is known.

\section{From HDR to CCG Surrogate}
\label{sec:hdr_model_sets}

This section constructs the CCG surrogate for each noise type.

\subsection{HDR as exact noise confidence region}

Given the noise density $f_W$ on $\mathbb{R}^q$ (with $q = nT$), the $(1-\alpha)$ HDR (Definition~\ref{def:hdr}) defines the exact noise confidence region:
\begin{equation}
\mathcal{H}_{W,1-\alpha} = \big\{W \in \mathbb{R}^{n \times T} : f_W(\mathrm{vec}(W)) \geq \tau_\alpha\big\},
\label{eq:hdr_noise}
\end{equation}
with $\Pr\{W_\star \in \mathcal{H}_{W,1-\alpha}\} = 1 - \alpha$.  Table~\ref{tab:hdr_examples} lists the HDR shapes considered.

\begin{table}[t]
\caption{HDR shape for the noise distributions considered.}
\label{tab:hdr_examples}
\centering
\footnotesize
\setlength{\tabcolsep}{2pt}
\renewcommand{\arraystretch}{0.98}
\begin{tabular*}{0.98\columnwidth}{@{\extracolsep{\fill}}p{0.25\columnwidth}p{0.46\columnwidth}c@{}}
\hline
Distribution & HDR shape & Convex \\
\hline
i.i.d.\ Gaussian & $\|W\|_F^2 \le \sigma^2\chi^2_{q,1-\alpha}$ & \checkmark \\
i.i.d.\ uniform & $\|W\|_\infty \le a$ & \checkmark \\
Gaussian mixture & non-convex, possibly disjoint & $\times$ \\
\hline
\end{tabular*}
\end{table}

\subsection{Exact likelihood-consistent model set}

The data equation $X_+ = \Theta M + W$ (Section~\ref{sec:data_equation}) defines the likelihood-consistent model set as all $\Theta$ whose residual lies in the HDR:
\begin{equation}
\mathcal{S}_{\Sigma,1-\alpha}^{\mathrm{exact}}
:= \big\{\Theta \in \mathbb{R}^{n \times (n+m)} : X_+ - \Theta M \in \mathcal{H}_{W,1-\alpha}\big\}.
\label{eq:exact_model_set}
\end{equation}
This set inherits the HDR geometry, with MLE $\hat{\Theta} = \arg\max_\Theta f_W(\mathrm{vec}(X_+ - \Theta M))$. Under bounded noise it reduces to the set-membership feasible set~\cite{amr23reachable}; under Gaussian noise, to a Frobenius ball; under Gaussian-mixture noise, to the corresponding non-convex geometry.

\begin{remark}[Scope]
\label{rem:nonconvex_barrier}
The developments below are exact for convex HDRs (bounded, Gaussian, mixed). Non-convex HDRs are treated via the MVEE surrogate in Section~\ref{sec:mvee_surrogate}.
\end{remark}

\subsection{CCG surrogate: definition and coverage guarantee}
\label{sec:ccg_surrogate}

\begin{definition}[CCG surrogate]
\label{def:ccg_surrogate}
A CCG surrogate for the $(1-\alpha)$ noise HDR $\mathcal{H}_{W,1-\alpha}$ is a CCG set (Definition~\ref{def:ccg})
\begin{equation}
\mathcal{E}_{W}^{\mathrm{CCG}}
= \left\{c_W + G_W \beta
\;\middle|\;
\|\beta_{\mathcal{I}_k}\|_{p_k} \le 1\ \forall\, k,\
A_0 \beta = b_0
\right\}
\label{eq:ccg_surrogate_def}
\end{equation}
satisfying $\mathcal{H}_{W,1-\alpha} \subseteq \mathcal{E}_{W}^{\mathrm{CCG}}$.
\end{definition}

\begin{proposition}[Coverage guarantee]
\label{prop:coverage}
If $\mathcal{H}_{W,1-\alpha} \subseteq \mathcal{E}_{W}^{\mathrm{CCG}}$, then
\begin{equation}
\Pr\{W_\star \in \mathcal{E}_{W}^{\mathrm{CCG}}\} \geq \Pr\{W_\star \in \mathcal{H}_{W,1-\alpha}\} = 1-\alpha.
\label{eq:coverage_guarantee}
\end{equation}
\end{proposition}

\subsection{Convex HDR: exact CCG representation}
\label{sec:direct_fit}

When the HDR is convex, the CCG surrogate can be constructed directly.

\noindent{\it Gaussian noise:} The HDR is the Frobenius ball $\|W\|_F^2 \le \sigma^2\chi^2_{q,1-\alpha}$. With $c_W = 0$, a single index group $p = 2$, and $G_W G_W^\top = \sigma^2\chi^2_{q,1-\alpha} I$, the CCG matches the HDR exactly as an ellipsotope.

\noindent{\it Bounded noise:}
The HDR is the box $\|W\|_\infty \le a$. With $c_W = 0$, singleton groups $p_k = \infty$, and $G_W = a I_{nT}$, the CCG coincides with the HDR in zonotopic form.

\begin{remark}[Approximation vs.\ exactness]
For Gaussian and bounded noise, the CCG surrogate is exact. For non-convex HDRs, such as Gaussian mixtures, a single convex CCG becomes approximate and must be interpreted as an outer surrogate.
\end{remark}

\subsection{Non-convex HDR: preliminary MVEE surrogate}
\label{sec:mvee_surrogate}

When the HDR is non-convex, a convex CCG cannot match it exactly. A simple remedy is to replace it by its minimum-volume enclosing ellipsoid.

\begin{proposition}[MVEE surrogate for non-convex HDRs]
\label{prop:mvee}
Let $\mathcal{H}_{W,1-\alpha} \subset \mathbb{R}^{q}$ be a bounded, possibly non-convex HDR, and let $\mathcal{E}_{W}^{\mathrm{MVEE}}$ denote its minimum-volume enclosing ellipsoid. Then
\begin{equation}
\mathcal{H}_{W,1-\alpha} \subseteq \mathcal{E}_{W}^{\mathrm{MVEE}},
\qquad
\Pr\{W_\star \in \mathcal{E}_{W}^{\mathrm{MVEE}}\} \ge 1-\alpha.
\label{eq:mvee_coverage}
\end{equation}
Moreover, $\mathcal{E}_{W}^{\mathrm{MVEE}}$ admits a one-group CCG representation with $p=2$.
\end{proposition}

\begin{compactproof}
By definition $\mathcal{E}_{W}^{\mathrm{MVEE}} \supseteq \mathcal{H}_{W,1-\alpha}$; coverage follows from Proposition~\ref{prop:coverage}. Any ellipsoid admits a one-group CCG with $p=2$.
\end{compactproof}

\begin{remark}[Limitation]
The MVEE preserves HDR coverage but is not exact. Richer representations such as polynomial CCG sets are left to future work.
\end{remark}

\section{Data-Consistent Constrained Matrix Convex Generators (CMCG)}
\label{sec:cmg_pullback}

This section derives the pullback from noise-level CCG to parameter-level CMCG, first in general and then for Gaussian, bounded, and mixed noise.

\subsection{Data equation}
\label{sec:data_equation}

Collect input--state data from~\eqref{eq:lti_system} into
\begin{align}
X_- &:= [x_0,\dots,x_{T-1}]\in\mathbb{R}^{n\times T},\nonumber\\
U_- &:= [u_0,\dots,u_{T-1}]\in\mathbb{R}^{m\times T},\nonumber\\
X_+ &:= [x_1,\dots,x_T]\in\mathbb{R}^{n\times T}.
\end{align}
Define $\Theta := \begin{bmatrix}A & B\end{bmatrix}\in\mathbb{R}^{n\times(n+m)}$ and $M := \begin{bmatrix}X_-\\ U_-\end{bmatrix}\in\mathbb{R}^{(n+m)\times T}$.
The data equation is
\begin{equation}
X_+ = \Theta M + W,
\label{eq:data_equation}
\end{equation}
with $W\in\mathbb{R}^{n\times T}$ stacking the disturbances.  We assume $M$ has full row rank.

\subsection{Pullback theorem: noise CCG to parameter CMCG}
\label{sec:pullback_theorem}

Let $\mathcal{E}_{W}^{\mathrm{CCG}}$ be a CCG surrogate as in~\eqref{eq:ccg_surrogate_def} with center $c_W$, generators $G_W^{(j)}$, index groups $\{\mathcal{I}_k, p_k\}$, and constraints $A_0 \beta = b_0$.  Let $M_\perp \in \mathbb{R}^{T \times d}$ span $\ker(M)$.

\begin{theorem}[Pullback]
\label{thm:pullback}
The data-consistent parameter set
\begin{equation}
\mathcal{N}_{\Sigma}^{\mathrm{CMCG}}
:= \big\{\Theta \in \mathbb{R}^{n \times (n+m)} : X_+ - \Theta M \in \mathcal{E}_{W}^{\mathrm{CCG}}\big\}
\end{equation}
is a CMCG (Definition~\ref{def:cmg}):
\begin{align}
\mathcal{N}_{\Sigma}^{\mathrm{CMCG}}
= \Big\{
&C_\Sigma + \textstyle\sum_j \beta_j G_\Sigma^{(j)}
\;\Big|\nonumber\\
&\|\beta_{\mathcal{I}_k}\|_{p_k} \le 1\ \forall\, k,\
A_c \beta = b_c
\Big\},
\label{eq:cmg_pullback}
\end{align}
with
\begin{equation}
C_\Sigma := (X_+ - c_W) M^\dagger, \quad
G_\Sigma^{(j)} := -G_W^{(j)} M^\dagger.
\label{eq:cmg_center_generators}
\end{equation}
The constraints combine CCG constraints with kernel solvability:
\begin{equation}
\begin{aligned}
A_c^{(j)} &:= \big[ A_0^{(j)\top}\ \ (G_W^{(j)} M_\perp)^\top \big]^\top, \\
b_c &:= \big[ b_0^\top\ \ ((X_+ - c_W) M_\perp)^\top \big]^\top.
\end{aligned}
\label{eq:cmg_constraints}
\end{equation}
Coverage carries over: $\Pr\{\Theta_\star \in \mathcal{N}_\Sigma^{\mathrm{CMCG}}\} \geq 1 - \alpha$.
\end{theorem}

\begin{compactproof}
The constraint $X_+ - \Theta M \in \mathcal{E}_{W}^{\mathrm{CCG}}$ means that there exists $\beta$ with $\|\beta_{\mathcal{I}_k}\|_{p_k} \le 1$ and $A_0 \beta = b_0$ such that $X_+ - \Theta M = c_W + \sum_j \beta_j G_W^{(j)}$.  For the linear equation $\Theta M = X_+ - c_W - \sum_j \beta_j G_W^{(j)}$ to be solvable in $\Theta$, the right-hand side must lie in the row space of $M$, i.e., $(X_+ - c_W - \sum_j \beta_j G_W^{(j)}) M_\perp = 0$.  This gives the kernel constraint $\sum_j \beta_j G_W^{(j)} M_\perp = (X_+ - c_W) M_\perp$.  The solution is then $\Theta = (X_+ - c_W - \sum_j \beta_j G_W^{(j)}) M^\dagger = C_\Sigma + \sum_j \beta_j G_\Sigma^{(j)}$.  The norm constraints on $\beta$ are inherited directly from the CCG surrogate, and the coverage follows from Proposition~\ref{prop:coverage}.
\end{compactproof}

The pullback is distribution-agnostic: the CMCG form depends only on the CCG surrogate; the noise model enters through $c_W$, $G_W^{(j)}$, and $\{\mathcal{I}_k, p_k\}$.

\subsection{Corollary 1: Gaussian noise --- const. matrix ellipsotope}
\label{sec:gaussian_cmg}

For i.i.d.\ Gaussian noise $W_{ij} \sim \mathcal{N}(0, \sigma^2)$, the HDR is the Frobenius ball $\|W\|_F^2 \le \sigma^2\chi^2_{q,1-\alpha}$ ($q = nT$), and the CCG surrogate is exact (Section~\ref{sec:direct_fit}) with $c_W = 0$ and a single $p = 2$ group.  A direct application of Theorem~\ref{thm:pullback} gives the intermediate set $\{\Theta \mid \|X_+ - \Theta M\|_F^2 \le \sigma^2\chi^2_{q,1-\alpha}\}$.  However, this $q$-dimensional ball is unnecessarily large: an orthogonal decomposition shows that only $d = n(n+m)$ of the $q$ noise dimensions affect the parameters, yielding the following tighter characterization.

\paragraph{Orthogonal decomposition.}
Let $\hat{\Theta} = X_+ M^\dagger$ be the OLS estimate, $P_M = M^\top(MM^\top)^{-1}M$ the projector onto the row space of~$M$.  Decompose $W = W_\parallel + W_\perp$ with $W_\parallel := W P_M$ and $W_\perp := W(I_T - P_M)$.  The estimation error depends only on $W_\parallel$:
\begin{equation}
\mathrm{tr}\!\big((\Theta - \hat{\Theta})\,MM^\top\,(\Theta - \hat{\Theta})^\top\big) = \|W_\parallel\|_F^2,
\label{eq:trace_identity_gauss}
\end{equation}
since $W_\perp M^\dagger = 0$.  Under the Gaussian assumption, $W_\parallel \perp\!\!\!\perp W_\perp$ and $\|W_\parallel\|_F^2/\sigma^2 \sim \chi^2_d$.  The $(1\!-\!\alpha)$ CMCG is therefore:
\begin{equation}
\mathcal{N}_\Sigma^{1-\alpha}
= \Big\{\Theta \mid
\mathrm{tr}\!\big((\Theta - \hat{\Theta})\, MM^\top\, (\Theta - \hat{\Theta})^\top\big)
\le \sigma^2\,\chi^2_{d,\,1-\alpha}
\Big\},
\label{eq:cmg_gauss}
\end{equation}
with coverage $\Pr\{\Theta_\star \in \mathcal{N}_\Sigma^{1-\alpha}\} = \Pr\{\|W_\parallel\|_F^2 \le \sigma^2\chi^2_{d,1-\alpha}\} = 1-\alpha$.  Note the radius uses $\chi^2_d$ ($d = n(n+m)$, the parameter dimension), not $\chi^2_q$ ($q = nT$, the noise dimension): the $q - d$ directions in $W_\perp$ do not influence the parameter estimate and are eliminated by the projection.

\paragraph{Equivalence with the MLE confidence ellipsoid.}
The estimation error $\hat{\Theta} - \Theta_\star = W M^\dagger$ is Gaussian with $\frac{1}{\sigma^2}\,\mathrm{tr}((\hat{\Theta} - \Theta_\star)\, MM^\top\, (\hat{\Theta} - \Theta_\star)^\top) \sim \chi^2_d$.  The $(1\!-\!\alpha)$ MLE confidence ellipsoid is
\begin{equation}
\mathcal{E}_\Theta^{1-\alpha}
= \Big\{\Theta \mid
\mathrm{tr}\!\big((\Theta - \hat{\Theta})\, MM^\top\, (\Theta - \hat{\Theta})^\top\big)
\le \sigma^2\,\chi^2_{d,1-\alpha}
\Big\}.
\label{eq:mle_ellipsoid}
\end{equation}
Comparing~\eqref{eq:cmg_gauss} and~\eqref{eq:mle_ellipsoid}, $\mathcal{N}_\Sigma^{1-\alpha} = \mathcal{E}_\Theta^{1-\alpha}$: the Gaussian CMCG coincides exactly with the MLE confidence ellipsoid.

\begin{proposition}[Containment hierarchy: CMCG $=$ MLE $\subseteq$ CMZ]
\label{prop:hierarchy}
For purely Gaussian noise, the CMCG~\eqref{eq:cmg_gauss} equals the MLE ellipsoid~\eqref{eq:mle_ellipsoid}.  Both are contained in the CMZ whenever the box $\|W\|_\infty \le m\sigma$ covers the $\chi^2_d$ ellipsoid.
\end{proposition}

\begin{remark}[Why the CMCG is much tighter than the CMZ]
\label{rem:mle_tighter}
The CMZ replaces the $\|\cdot\|_2$-ball by a $\|\cdot\|_\infty$-box in all $q = nT$ noise coordinates, with volume inflation exponential in $q$ (Remark~\ref{rem:norm_mismatch}).  The CMCG uses $\chi^2_d$ ($d = n(n\!+\!m) \ll q$) because the remaining $q\!-\!d$ directions do not affect parameters. For $n\!=\!1$, $T\!=\!30$: CMZ operates in $q\!=\!30$ dimensions, CMCG in $d\!=\!2$.
\end{remark}

\begin{table}[t]
\caption{Structural parallel between the bounded and Gaussian noise frameworks.}
\label{tab:parallel}
\centering
\small
\setlength{\tabcolsep}{3pt}
\renewcommand{\arraystretch}{1.05}
\begin{tabular*}{\columnwidth}{@{\extracolsep{\fill}}lcc@{}}
\hline
Level & \shortstack{Bounded\\($\|\cdot\|_\infty$)} & \shortstack{Gaussian\\($\|\cdot\|_2$)} \\
\hline
Noise set & zonotope & ellipsoid \\
Unconstr.\ & MZ & ME \\
+ kernel & CMZ (exact) & CMCG $=$ MLE \\
\hline
\end{tabular*}
\end{table}

\subsection{Corollary 2: Bounded-support noise --- CMZ}
\label{sec:bounded_cmg}

Suppose each entry of $W$ is bounded: $|W_{ij}| \le a$.  The HDR is the box $\|W\|_\infty \le a$, which the CCG exactly represents as a zonotope ($p = \infty$, singleton index groups).  The pullback (Theorem~\ref{thm:pullback}) yields a constrained matrix zonotope (CMZ).

The CMZ form was established in~\cite{amr23reachable}; we restate it for completeness.  The noise set is $\mathcal{M}_w = \{C_w + \sum_i \beta_i G_w^{(i)} \mid \|\beta\|_\infty \le 1\}$, and kernel solvability $(X_+ - W)M_\perp = 0$ yields
\begin{gather}
\textstyle\sum_{i=1}^{\gamma_w}\beta_i A_w^{(i)} = B_w, \nonumber\\
A_w^{(i)} := G_w^{(i)}M_\perp,\quad
B_w := (X_+\!-\!C_w)M_\perp.
\label{eq:Nw_constraint_matrices}
\end{gather}
The parameter set is the CMZ
\begin{equation}
\mathcal{N}_\Sigma
=
\left\{
C_\Sigma + \textstyle\sum_{i=1}^{\gamma_w}\beta_i G_\Sigma^{(i)}
\;\middle|\;
\textstyle\sum_{i}\beta_i A_w^{(i)} \!=\! B_w,\; \|\beta\|_\infty\!\le\! 1
\right\}\!,
\label{eq:final_cmz}
\end{equation}
with $C_\Sigma = (X_+ - C_w)M^\dagger$ and $G_\Sigma^{(i)} = -G_w^{(i)} M^\dagger$.

\paragraph{MLE equivalence under uniform noise.}
When the noise is i.i.d.\ uniform, $W_{ij} \sim \mathrm{Unif}([-a,a])$, the likelihood is flat over its support:
\begin{equation}
L(\Theta)=(2a)^{-nT}\,\mathbf 1\!\left(\|X_+-\Theta M\|_\infty\le a\right).
\label{eq:uniform_likelihood}
\end{equation}
Every feasible $\Theta$ maximizes $L$, so the MLE solution set equals the feasible model set:
\begin{equation}
\arg\max_\Theta L(\Theta)
=\{\Theta \mid \|X_+-\Theta M\|_\infty\le a\}
=\mathcal N_\Sigma.
\label{eq:mle_equals_setmembership}
\end{equation}
Under uniform noise, set-membership identification coincides with maximum-likelihood estimation~\cite{knight2020linfty,yi2024linfty}.

\subsection{Mixed bounded-Gaussian noise}
\label{sec:mixed_noise_cmg}

Consider the additive mixed noise model
\begin{equation}
w_k = w_{b,k} + w_{g,k},
\label{eq:mixed_noise_additive}
\end{equation}
where $w_{b,k} = G_b\,\beta_k$ with $\|\beta_k\|_\infty \le 1$ and $G_b \in \mathbb{R}^{n \times p_b}$, and $w_{g,k} \sim \mathcal{N}(0, \sigma^2 I_n)$, independent across $k$.

\paragraph{Mixed-index noise confidence region.}
Stacking over $T$ steps, $W = W_b + W_g$ with $W_b$ a matrix zonotope ($|\beta_k^{(b)}| \le 1$) and $W_g$ Gaussian.  At the noise level, the Gaussian part is truncated by its $q$-dimensional Frobenius ball ($\|\beta^{(g)}\|_2 \le 1$).  The noise confidence region is a mixed-index CCG:
\begin{equation}
\mathcal{W}_{1-\alpha}^{\mathrm{mix}}
= \left\{
\sum_{k} \beta_k^{(b)} G_{W_b}^{(k)}
+ \sum_{k} \beta_k^{(g)} G_{W_g}^{(k)}
\;\middle|\;
\begin{aligned}
&\|\beta^{(b)}\|_\infty \le 1, \\
&\|\beta^{(g)}\|_2 \le 1
\end{aligned}
\right\}.
\label{eq:mixed_noise_confidence}
\end{equation}

\paragraph{The CMCG for mixed noise.}
Applying Theorem~\ref{thm:pullback} to~\eqref{eq:mixed_noise_confidence} gives:
\begin{align}
C_\Sigma &:= X_+ M^\dagger, \quad
G_{\Sigma,b}^{(k)} := -G_{W_b}^{(k)} M^\dagger, \nonumber\\
G_{\Sigma,g}^{(k)} &:= -G_{W_g}^{(k)} M^\dagger, \quad
A_b^{(k)} := G_{W_b}^{(k)} M_\perp, \nonumber\\
A_g^{(k)} &:= G_{W_g}^{(k)} M_\perp, \quad
B_W := X_+ M_\perp.
\label{eq:cmg_mixed_pieces}
\end{align}
The parameter set is the CMCG:
\begin{align}
\mathcal{N}_\Sigma^{1-\alpha}
= \Big\{
&C_\Sigma + \textstyle\sum_k \beta_k^{(b)} G_{\Sigma,b}^{(k)}
+ \textstyle\sum_k \beta_k^{(g)} G_{\Sigma,g}^{(k)}
\;\Big|\nonumber\\
&\|\beta^{(b)}\|_\infty \!\le\! 1,\;
\|\beta^{(g)}\|_2 \!\le\! 1,\nonumber\\
&\textstyle\sum_k \beta_k^{(b)} A_b^{(k)} + \sum_k \beta_k^{(g)} A_g^{(k)} = B_W
\Big\}.
\label{eq:cmg_mixed}
\end{align}

\begin{proposition}[Coverage of the mixed CMCG with $\chi^2_d$ radius]
\label{prop:mixed_coverage}
In the CMCG~\eqref{eq:cmg_mixed}, the Gaussian generators $G_{\Sigma,g}^{(k)}$ are scaled by $r_g = \sigma\sqrt{\chi^2_{d,1-\alpha}}$ with $d = n(n+m)$.  Then $\Pr\{\Theta_\star \in \mathcal{N}_\Sigma^{1-\alpha}\} \ge 1 - \alpha$.
\end{proposition}

\begin{compactproof}
Three facts are used: (i)~$W_b$ and $W_g$ are independent by the noise model~\eqref{eq:mixed_noise_additive}; (ii)~$W_{g,\parallel} := W_g P_M$ and $W_{g,\perp} := W_g(I_T - P_M)$ are independent under Gaussianity, and $\|W_{g,\parallel}\|_F^2/\sigma^2 \sim \chi^2_d$ (Section~\ref{sec:gaussian_cmg}); (iii)~the parameter estimate depends on $W_g$ only through $W_{g,\parallel}$ (since $W_{g,\perp} M^\dagger = 0$), so the $\chi^2_d$ distribution of $W_{g,\parallel}$ is not affected by the presence of $W_b$.  Therefore
\begin{multline*}
\Pr\{\Theta_\star \in \mathcal{N}_\Sigma^{1-\alpha}\}
= \underbrace{\Pr\{W_b \in \mathcal{M}_{W_b}\}}_{=\,1}\\
\times\; \underbrace{\Pr\{\|W_{g,\parallel}\|_F^2 \le \sigma^2\chi^2_{d,1-\alpha}\}}_{=\,1-\alpha}
= 1 - \alpha. \qedhere
\end{multline*}
\end{compactproof}

\begin{remark}[CMCG as bridge]
\label{rem:bridge}
The CMCG~\eqref{eq:cmg_mixed} unifies the noise scenarios: $\sigma = 0$ recovers the CMZ~\eqref{eq:final_cmz}; $G_b = 0$ recovers the MLE ellipsoid~\eqref{eq:cmg_gauss}. In the mixed case, both generator families remain present: the bounded generators with $\|\beta^{(b)}\|_\infty \le 1$ capture worst-case set-membership uncertainty, while the Gaussian generators with $\|\beta^{(g)}\|_2 \le 1$ and radius $r_g = \sigma\sqrt{\chi^2_{d,1-\alpha}}$ retain the exact ellipsoidal confidence geometry at the parameter level.
\end{remark}

\begin{proposition}[Tightness over CMZ]
\label{prop:tighter_than_cmz}
Let $\mathcal{N}_\Sigma^{\mathrm{CMZ}}$ denote the CMZ obtained by replacing the Gaussian noise $w_{g,k}$ by the box $\|w_{g,k}\|_\infty \le m\sigma$ (e.g., $m = 3$)~\cite{amr23reachable}.  Then $\mathcal{N}_\Sigma^{1-\alpha} \subseteq \mathcal{N}_\Sigma^{\mathrm{CMZ}}$, with the inclusion strict whenever $q_g \ge 2$.
\end{proposition}

\begin{compactproof}
The box $\|w_{g,k}\|_\infty \le m\sigma$ contains the $2$-norm ball, strictly so for $n \ge 2$ since $2^n / V_n > 1$.  This noise-level inclusion propagates to the parameter level via $W \mapsto (X_+ - W)M^\dagger$.
\end{compactproof}

\begin{remark}[Optimality of the mixed CMCG]
\label{rem:perp_cannot_help_bounded}
One might try to use the distribution of $W_{g,\perp}$ to further tighten the bounded coefficients via $W_{b,\perp} = R_\perp - W_{g,\perp}$.  However, bounded generators are typically low-rank (rank-$1$ when $p_b = 1$), making the resulting LP infeasible, and the coupling $\|W_b\|_\infty \le a$ between $W_{b,\parallel}$ and $W_{b,\perp}$ prevents the orthogonal independence needed for further projection.  The CMCG is thus equivalent to a profile likelihood approach: bounded noise is handled by set-membership, Gaussian noise by its marginal likelihood over the parameter-identifiable subspace, using $\chi^2_d$ rather than $\chi^2_q$.  Their Minkowski-sum combination is already the tightest achievable for mixed noise.
\end{remark}

\section{Forward Propagation and Numerical Evaluation}
\label{sec:forward_prob_sets}

The CMCG $\times$ CCG multiplication preserves the correct $p$-norm for each generator type ($2$-norm for Gaussian, $\infty$-norm for bounded), whereas standard zonotope propagation treats all generators with $\|\cdot\|_\infty$.

\subsection{Basic CCG operations}
Let $\mathcal{E}_1$ and $\mathcal{E}_2$ be two CCG sets of the form~\eqref{eq:ccg_def}, with bounded generators $G_{b,i}$, Gaussian generators $G_{g,i}$, and constraints $A_{b,i}$, $A_{g,i}$, $B_i$ for $i = 1, 2$. Then
\begin{align}
&\mathcal{E}_1 \oplus \mathcal{E}_2
= \big\{c_1+c_2 + G_{b,1}\beta_1^{(b)} + G_{b,2}\beta_2^{(b)}
\nonumber\\
&\quad + G_{g,1}\beta_1^{(g)} + G_{g,2}\beta_2^{(g)}
\ \big|\
\|\beta_i^{(b)}\|_\infty\!\le\!1,\
\|\beta_i^{(g)}\|_2\!\le\!1\big\},
\nonumber\\
&A_b\!:=\!\mathrm{blkdiag}(A_{b,1},A_{b,2}),\
A_g\!:=\!\mathrm{blkdiag}(A_{g,1},A_{g,2}),
\nonumber\\
&B\!:=\!\big[B_1^\top\ B_2^\top\big]^\top,
\label{eq:mie_minkowski}
\end{align}
and for any linear map $R$,
\begin{equation}
R\mathcal{E}_1
= \langle Rc_1,\ RG_{b,1},\ RG_{g,1},\ A_{b,1},\ A_{g,1},\ B_1\rangle.
\label{eq:mie_linear_map}
\end{equation}
Both norm and equality constraints are preserved through block-diagonal augmentation.

\subsection{Multiplying a CMCG by a CCG}
Let $\Theta \in \mathcal{E}_\Theta$ be a CMCG as in~\eqref{eq:cmg_mixed} and
let $z$ be a CCG with center $c_z$, bounded generators $G_{z,b}^{(\ell)}$, and Gaussian generators $G_{z,g}^{(r)}$.
The product $y = \Theta z$ is over-approximated by a CCG with the following components:
\begin{align}
c_y &= C_\Sigma\,c_z, \label{eq:prod_center}\\
G_{y,b} &= \big[\,
\underbrace{G_{\Sigma,b}^{(k)}\!c_z}_{\scriptscriptstyle A_b}\;\big|\;
\underbrace{C_\Sigma G_{z,b}}_{\scriptscriptstyle\text{lin}}\;\big|\;
\underbrace{d_{k\ell}\,G_{\Sigma,b}^{(k)}\!G_{z,b}^{(\ell)}}_{\scriptscriptstyle\text{b}\!\times\!\text{b}}
\nonumber\\
&\qquad\;\big|\;
\underbrace{\rho_\Theta\rho_z\,G_{\Sigma,g}^{(j)}\!G_{z,g}^{(r)}}_{\scriptscriptstyle\text{e}\!\times\!\text{e}}
\,\big], \label{eq:prod_Gb}\\
G_{y,g} &= \big[\,
\underbrace{G_{\Sigma,g}^{(j)}\!c_z}_{\scriptscriptstyle A_g}\;\big|\;
\underbrace{C_\Sigma G_{z,g}}_{\scriptscriptstyle\text{lin}}\;\big|\;
\underbrace{\bar{\beta}_k\,G_{\Sigma,b}^{(k)}\!G_{z,g}^{(r)}}_{\scriptscriptstyle\text{b}\!\times\!\text{e}}
\nonumber\\
&\qquad\;\big|\;
\underbrace{\bar{\alpha}_\ell\,G_{\Sigma,g}^{(j)}\!G_{z,b}^{(\ell)}}_{\scriptscriptstyle\text{e}\!\times\!\text{b}}
\,\big], \label{eq:prod_Gg}
\end{align}
Here $d_{k\ell} = \bar{\beta}_k \bar{\alpha}_\ell$, where $\bar{\beta}_k$ and $\bar{\alpha}_\ell$ are upper bounds on $|\beta_k^{(b)}|$ and $|\alpha_\ell^{(b)}|$ from the $\|\cdot\|_\infty$ constraints or auxiliary LPs. The radii $\rho_\Theta = \sqrt{\chi^2_{\gamma_{g,\Theta},\, 1-\delta/2}}$ and $\rho_z = \sqrt{\chi^2_{\gamma_{g,z},\, 1-\delta/2}}$ truncate the Gaussian coefficients into a confidence event with probability $\ge 1-\delta$, converting the Gaussian$\times$Gaussian bilinear term into a bounded block. Unlike prior probabilistic zonotope constructions, these radii come from the $\chi^2$ distribution of $\|\xi\|_2^2$ rather than a box $\|\xi\|_\infty \le m$, avoiding the volume inflation of Remark~\ref{rem:norm_mismatch}.

The constraint matrices are padded with zeros so that only the original coefficients remain coupled:
\begin{equation}
A_b^{\mathrm{out}} = \big[A_b\ \ 0\big],\qquad
A_g^{\mathrm{out}}   = \big[A_g\ \ 0\big],\qquad
B^{\mathrm{out}} = B_W.
\label{eq:prod_constraints}
\end{equation}

The first two blocks of \eqref{eq:prod_Gb}--\eqref{eq:prod_Gg} retain the original coefficients; the bilinear blocks introduce fresh variables: $\delta^{bb}$ ($\|\cdot\|_\infty \le 1$), $\eta_k^{bg}$ and $\eta_\ell^{gb}$ ($\|\cdot\|_2 \le 1$), and $\lambda^{gg}$ ($\|\cdot\|_\infty \le 1$). The equality constraints~\eqref{eq:prod_constraints} act only on the original coefficients.

\begin{theorem}[Containment of the CMCG $\times$ CCG over-approximation]
\label{thm:product_containment}
Let $\mathcal{P}_{\delta}(\mathcal{E}_\Theta,\mathcal{E}_z)$ denote the exact product set $\{\Theta z\}$ generated by all admissible bounded coefficients and by all Gaussian coefficients satisfying the confidence event $\|\xi_\Theta\|_2 \le \rho_\Theta$, $\|\xi_z\|_2 \le \rho_z$. Then
\begin{equation}
\mathcal{P}_{\delta}(\mathcal{E}_\Theta,\mathcal{E}_z)
\subseteq
\mathcal{E}_y,
\label{eq:product_containment}
\end{equation}
where $\mathcal{E}_y$ is the CCG defined by \eqref{eq:prod_center}--\eqref{eq:prod_constraints}.
\end{theorem}

\begin{compactproof}
Write
\begin{gather*}
\Theta = C_\Sigma
+ \textstyle\sum_k \beta_k^{(b)} G_{\Sigma,b}^{(k)}
+ \textstyle\sum_j \xi_{\Theta,j} G_{\Sigma,g}^{(j)},\\
z = c_z
+ \textstyle\sum_\ell \alpha_\ell^{(b)} G_{z,b}^{(\ell)}
+ \textstyle\sum_r \xi_{z,r} G_{z,g}^{(r)}.
\end{gather*}
Expanding $y=\Theta z$ gives eight groups of terms: center$\times$center, two center$\times$generator blocks, two generator$\times$center blocks, and four bilinear blocks. The linear blocks are represented exactly by the first two blocks of \eqref{eq:prod_Gb} and \eqref{eq:prod_Gg}, while the original equality constraints on $\beta^{(b)}$ and $\xi_\Theta$ are retained through \eqref{eq:prod_constraints}. For the bounded$\times$bounded block,
\[
\beta_k^{(b)} \alpha_\ell^{(b)}
=
\delta_{k\ell}^{bb}\,\bar{\beta}_k \bar{\alpha}_\ell,
\qquad
|\delta_{k\ell}^{bb}| \le 1,
\]
so the term is contained in the generator block $d_{k\ell} G_{\Sigma,b}^{(k)} G_{z,b}^{(\ell)}$. For the bounded$\times$Gaussian block, define $\eta_k^{bg} := (\beta_k^{(b)}/\bar{\beta}_k)\,\xi_z$; then $\|\eta_k^{bg}\|_2 \le 1$, so the corresponding term lies in the block $\bar{\beta}_k G_{\Sigma,b}^{(k)} G_{z,g}^{(r)}$. Similarly, for the Gaussian$\times$bounded block, $\eta_\ell^{gb} := (\alpha_\ell^{(b)}/\bar{\alpha}_\ell)\,\xi_\Theta$ satisfies $\|\eta_\ell^{gb}\|_2 \le 1$, so the term lies in the block $\bar{\alpha}_\ell G_{\Sigma,g}^{(j)} G_{z,b}^{(\ell)}$. Finally, on the event $\|\xi_\Theta\|_2 \le \rho_\Theta$, $\|\xi_z\|_2 \le \rho_z$, each coefficient product satisfies $|\xi_{\Theta,j}\xi_{z,r}| \le \rho_\Theta \rho_z$. Hence
\[
\xi_{\Theta,j}\xi_{z,r}
=
\lambda_{jr}^{gg}\,\rho_\Theta \rho_z,
\qquad
|\lambda_{jr}^{gg}| \le 1,
\]
which places the Gaussian$\times$Gaussian block in the bounded generator family of \eqref{eq:prod_Gb}. Therefore every exact product realization belongs to $\mathcal{E}_y$.
\end{compactproof}

\begin{remark}[Where the over-approximation enters]
\label{rem:where_overapprox_enters}
Linear terms are exact. Over-approximation enters in the bilinear blocks, where products of shared coefficients are replaced by fresh variables, dropping algebraic dependence (the wrapping effect). The Gaussian$\times$Gaussian block adds further conservatism by replacing the rank-one matrix $\xi_\Theta \xi_z^\top$ with independent bounded coefficients $\lambda_{jr}^{gg}$.
\end{remark}

\begin{proposition}[Rough bound for the Gaussian$\times$Gaussian block]
\label{prop:gg_gap}
Let
\[
H_{jr} := G_{\Sigma,g}^{(j)} G_{z,g}^{(r)},
\]
and define the exact truncated Gaussian$\times$Gaussian set
\[
\mathcal{S}_{gg}
:=
\left\{
\sum_{j,r} \xi_{\Theta,j}\xi_{z,r} H_{jr}
\ \middle|\
\|\xi_\Theta\|_2 \le \rho_\Theta,\;
\|\xi_z\|_2 \le \rho_z
\right\},
\]
and its bounded-generator over-approximation
\[
\widehat{\mathcal{S}}_{gg}
:=
\left\{
\sum_{j,r} \lambda_{jr} \rho_\Theta \rho_z H_{jr}
\ \middle|\
\|\lambda\|_\infty \le 1
\right\}.
\]
For any support direction $h$ with $\|h\|_2 = 1$,
\begin{multline}
0 \le h_{\widehat{\mathcal{S}}_{gg}}(h) - h_{\mathcal{S}_{gg}}(h)\\
\le
\rho_\Theta \rho_z \big(\!\sqrt{\gamma_{g,\Theta}\gamma_{g,z}}-1\big)
\Big(\sum_{j,r}\|H_{jr}\|_F^2\Big)^{\!1/2}\!.
\label{eq:gg_gap_bound}
\end{multline}
\end{proposition}

\begin{compactproof}
Let $M_h \in \mathbb{R}^{\gamma_{g,\Theta}\times\gamma_{g,z}}$ be defined by $(M_h)_{jr} := \langle h, H_{jr}\rangle$. Then
\[
h_{\mathcal{S}_{gg}}(h)
=
\rho_\Theta \rho_z \|M_h\|_2,
\qquad
h_{\widehat{\mathcal{S}}_{gg}}(h)
=
\rho_\Theta \rho_z \sum_{j,r} |(M_h)_{jr}|.
\]
Therefore
\[
0 \le h_{\widehat{\mathcal{S}}_{gg}}(h) - h_{\mathcal{S}_{gg}}(h)
\le
\rho_\Theta \rho_z \big(\|M_h\|_{1,\mathrm{entry}} - \|M_h\|_2\big).
\]
Using $\|M_h\|_{1,\mathrm{entry}} \le \sqrt{\gamma_{g,\Theta}\gamma_{g,z}}\,\|M_h\|_F$ and $\|M_h\|_2 \le \|M_h\|_F$ gives
\[
\|M_h\|_{1,\mathrm{entry}} - \|M_h\|_2
\le
\big(\sqrt{\gamma_{g,\Theta}\gamma_{g,z}}-1\big)\|M_h\|_F.
\]
Finally,
\[
\|M_h\|_F^2
=
\sum_{j,r} \langle h, H_{jr}\rangle^2
\le
\sum_{j,r} \|H_{jr}\|_F^2 \|h\|_2^2,
\]
which yields \eqref{eq:gg_gap_bound}.
\end{compactproof}

\begin{remark}[Wrapping error propagation]
\label{rem:wrapping_recurrence}
With $\kappa := \sup_{\Theta \in \mathcal{N}_\Sigma^{1-\alpha}} \|\Theta\|_2$ and one-step error $\varepsilon_{\mathrm{prod}}$ (bounded by Proposition~\ref{prop:gg_gap}), the Hausdorff error satisfies $d_H(\widetilde{\mathcal{R}}_{k+1}, \mathcal{R}_{k+1}^{\mathrm{exact}}) \le \kappa\, d_H(\widetilde{\mathcal{R}}_{k}, \mathcal{R}_{k}^{\mathrm{exact}}) + \varepsilon_{\mathrm{prod}}$, giving $d_H \le \frac{1-\kappa^K}{1-\kappa}\varepsilon_{\mathrm{prod}}$ for $\kappa < 1$.  Since $\kappa$ is the same for both schemes, the $2$-norm improvement is not washed out.
\end{remark}

\subsection{End-to-end one-step reachability}
For the system $x_{k+1} = Ax_k + Bu_k + w_{b,k} + w_{g,k}$ with
$\Theta = [A\ B] \in \mathcal{N}_\Sigma^{1-\alpha}$ (CMCG),
$w_{b,k} \in \mathcal{W}_b$ (zonotopic), and $w_{g,k}$ with $\|w_{g,k}\|_2 \le \sigma\sqrt{\chi^2_{n,1-\alpha_w}}$ (ellipsoidal), define the augmented state-input set $\mathcal{Z}_k := \mathcal{X}_k \times \mathcal{U}$.  The one-step reachable set satisfies
\begin{equation}
\mathcal{X}_{k+1} \supseteq
\mathcal{N}_\Sigma^{1-\alpha} \times \mathcal{Z}_k
\ \oplus\ \mathcal{W}_b\ \oplus\ \mathcal{W}_g,
\label{eq:reach_step}
\end{equation}
with $\times$ the CMCG--CCG product, $\mathcal{Z}_k = \mathcal{X}_k \times \mathcal{U}$, and $\oplus$ the Minkowski sum.  By \eqref{eq:mie_minkowski}--\eqref{eq:prod_constraints}, each term remains a CCG, preserving the bounded/Gaussian distinction.

\begin{proposition}[Guaranteed outer bound]
\label{prop:outer_bound}
At every propagation step $k$, the CMCG-based reachable set satisfies $\mathcal{R}_k^{\mathrm{true}} \subseteq \mathcal{R}_k^{\mathrm{CMCG}}$, where $\mathcal{R}_k^{\mathrm{true}}$ is the true reachable set under all admissible noise realizations and system matrices in $\mathcal{S}_{\Sigma,1-\alpha}^{\mathrm{exact}}$.
\end{proposition}
This follows from $\mathcal{S}_{\Sigma,1-\alpha}^{\mathrm{exact}} \subseteq \mathcal{N}_\Sigma^{\mathrm{CMCG}}$, Theorem~\ref{thm:product_containment}, and set-monotonicity.

\subsection{Numerical evaluation}
\label{sec:evaluation}

We validate the proposed approach with three numerical studies. The first compares the parameter sets produced by CMCG, MLE, and CMZ. The second compares CMCG-based and CMZ-based reachability~\cite{amr23reachable}. The third illustrates a preliminary Gaussian-mixture treatment through the MVEE surrogate introduced in Proposition~\ref{prop:mvee}.

\subsubsection{Experiment~1: Parameter-Set Hierarchy}

Consider a scalar system $x_{k+1} = a\,x_k + b\,u_k + w_k$ with $n=1$, $m=1$, $T=30$, $\sigma = 0.02$, and confidence level $1 - \alpha = 0.95$.  The noise dimension is $q = nT = 30$ and the parameter dimension $d = n(n+m) = 2$.

\begin{figure}[t]
    \centering
    \includegraphics[width=0.92\columnwidth]{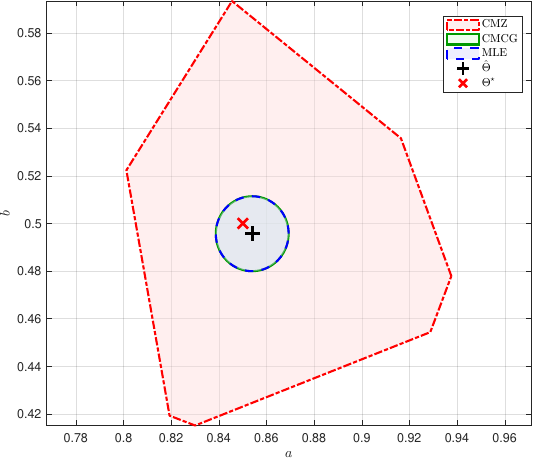}
    \caption{Parameter-set comparison for a scalar system ($n\!=\!1$, $T\!=\!30$).  The CMZ (red, dash-dot) over-approximates Gaussian noise by a $5\sigma$ box in $q\!=\!30$ dimensions, yielding a large polytope.  The CMCG (green, solid) uses the $\chi^2_d$ radius with $d\!=\!2$, coinciding exactly with the MLE ellipsoid (blue, dashed).  OLS estimate $\hat{\Theta}$ (black~$+$); true parameters $\Theta^\star$ (red~$\times$).}
    \label{fig:param_ellipsoids}
\end{figure}

Fig.~\ref{fig:param_ellipsoids} shows the parameter sets in the $(a,b)$-plane.  The CMCG (green) and MLE (blue, dashed) coincide exactly, confirming Proposition~\ref{prop:hierarchy}.  The CMZ (red) is much larger because it replaces the $2$-norm ball by a $5\sigma$ box in $q=30$ dimensions (Remark~\ref{rem:norm_mismatch}), while the CMCG uses only the $d=2$ parameter-relevant directions.

\subsubsection{Experiment~2: CMCG vs.\ CMZ Reachability}

We compare CMCG-based and CMZ-based~\cite{amr23reachable} reachability on a $5$-dimensional system ($n = 5$, $m = 1$, $\Delta t = 0.05$\,s) with mixed noise $w_k = w_{b,k} + w_{g,k}$, $w_{b,k} \in [-a, a]^n$ ($a = 10^{-4}$), $w_{g,k} \sim \mathcal{N}(0, \sigma^2 I_n)$ ($\sigma = 6\times 10^{-4}$), and $T = 120$ samples. The Gaussian component is six times larger than the bounded one.

\begin{figure*}[!t]
    \centering
    \includegraphics[width=\textwidth]{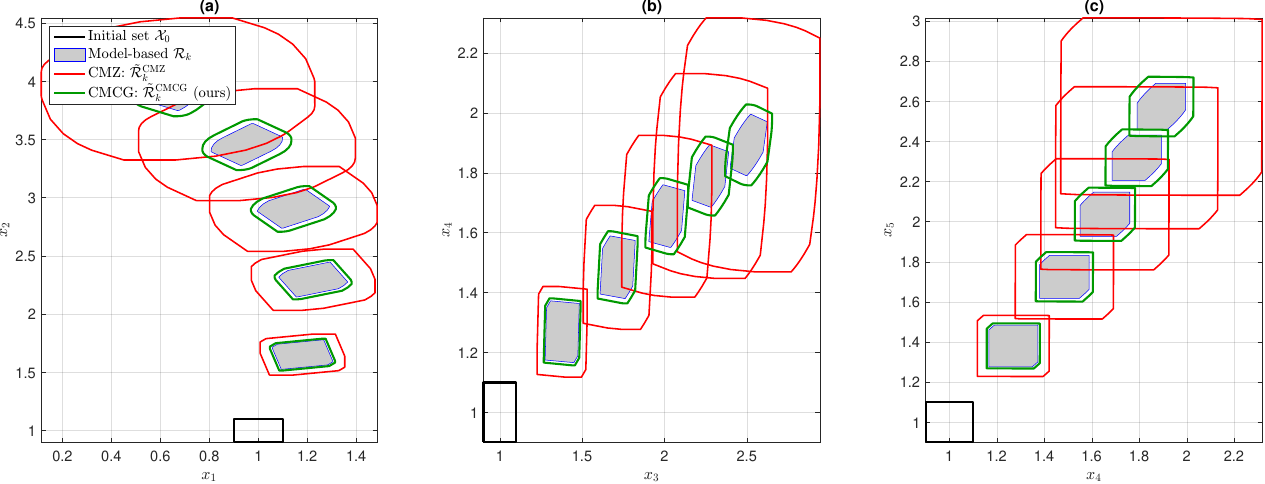}
    \caption{Reachable-set comparison over 5 propagation steps for a 5D system, shown in three 2D projections: (a)~$(x_1, x_2)$, (b)~$(x_3, x_4)$, (c)~$(x_4, x_5)$.  $\mathcal{R}_k$ denotes the model-based reachable set (blue, gray fill), $\tilde{\mathcal{R}}_k^{\mathrm{CMZ}}$ the CMZ over-approximation (red, outermost), and $\tilde{\mathcal{R}}_k^{\mathrm{CMCG}}$ our CMCG-based set (green).  The CMCG sets are consistently tighter because the CCG propagation preserves the correct $2$-norm for Gaussian generators.}
    \label{fig:cmg_vs_cmz}
\end{figure*}

Fig.~\ref{fig:cmg_vs_cmz} shows five propagation steps. The hierarchy $\mathcal{R}_k \subseteq \tilde{\mathcal{R}}_k^{\mathrm{CMCG}} \subseteq \tilde{\mathcal{R}}_k^{\mathrm{CMZ}}$ holds at every step, with the gap widening with dimension as predicted by the volume ratio in Remark~\ref{rem:norm_mismatch}.

Table~\ref{tab:exp2_comparison} quantifies the gap: $V_{\mathrm{CMZ}}/V_{\mathrm{CMCG}} = 221.6\times$ at $k=5$, and the CMCG is $1275\times$ faster because the CCG product avoids the LP solves of the kernel-constrained CMZ.

\begin{table}[t]
\caption{Computation time and final interval-hull volume for the 5D reachability problem ($T=120$, $K=5$ steps).}
\label{tab:exp2_comparison}
\centering
\small
\setlength{\tabcolsep}{4pt}
\renewcommand{\arraystretch}{1.1}
\begin{tabular*}{\columnwidth}{@{\extracolsep{\fill}}l rrr@{}}
\hline
 & Model & CMZ & CMCG \\
\hline
Offline time (s)  & $<$0.01 & 274.1 & 0.13 \\
\textbf{Total time (s)} & \textbf{0.01} & \textbf{275.8} & \textbf{0.20} \\
Final volume ($k\!=\!5$) & 1.12e-3 & 8.85e-1 & 3.99e-3 \\
\hline
\end{tabular*}
\end{table}

\subsubsection{Experiment~3: Gaussian-Mixture Noise via MVEE}

Consider the same scalar system but with bimodal noise
\begin{equation}
w_k \sim \tfrac{1}{2}\mathcal{N}(-\mu,\sigma^2) + \tfrac{1}{2}\mathcal{N}(\mu,\sigma^2),
\qquad
\mu = 0.15,\ \sigma = 0.05.
\label{eq:gmm_noise}
\end{equation}
The marginal HDR splits into two disjoint intervals; we replace it by the MVEE (Proposition~\ref{prop:mvee}) and propagate the resulting CMCG for five steps.

\begin{figure}[t]
    \centering
    \includegraphics[width=0.92\columnwidth]{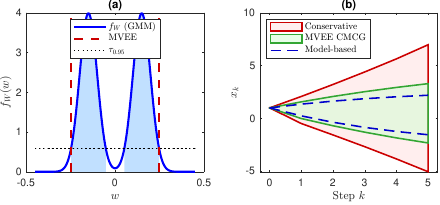}
    \caption{Gaussian-mixture case study. (a)~Bimodal scalar density with its $95\%$ HDR (shaded) and the MVEE surrogate (red dashed). (b)~Five-step reachable sets: conservative single-Gaussian (red), MVEE-based CMCG (green), and model-based (blue dashed).  The MVEE construction gives a tighter outer approximation by convexifying the non-convex HDR.}
    \label{fig:gmm_experiment}
\end{figure}

Fig.~\ref{fig:gmm_experiment} shows the bimodal density with its HDR and MVEE surrogate~(a), and the five-step reachable sets~(b). The MVEE-based CMCG remains a valid outer approximation while being substantially tighter than a conservative single-Gaussian surrogate.

\subsubsection{Discussion}

All three experiments confirm the theory: CMCG $=$ MLE $\subset$ CMZ for Gaussian noise (Fig.~\ref{fig:param_ellipsoids}), $\tilde{\mathcal{R}}_k^{\mathrm{CMCG}} \subset \tilde{\mathcal{R}}_k^{\mathrm{CMZ}}$ at every step with a volume ratio of $221.6\times$ at $k=5$ (Table~\ref{tab:exp2_comparison}), and the MVEE surrogate handles non-convex noise (Fig.~\ref{fig:gmm_experiment}).

\section{Conclusion}\label{sec:conclusion}
This paper shows how mixed-$p$ CCG/CMCG sets systematically improve data-driven reachability by keeping the correct norm for each noise component. The CMCG coincides with the MLE ellipsoid for Gaussian noise ($\text{CMCG} = \text{MLE} \subset \text{CMZ}$) and remains strictly tighter than the CMZ for mixed bounded-Gaussian noise, with a formal containment proof for the CMCG $\times$ CCG product. As a result, the proposed approach yields substantially tighter and less conservative reachable sets while maintaining computational tractability. Numerical results confirm both improved accuracy and efficiency in reachable-set computation. These properties make the approach particularly relevant for safety verification and uncertainty-aware control design in data-driven settings. Future work includes polynomial CCG sets for exact non-convex HDR representations and conformal prediction~\cite{csaji2012sps} for distribution-free guarantees.

\end{document}